\documentclass[epj,nopacs]{svjour}

\pdfoutput=1
\usepackage{lmodern}
\usepackage[T1]{fontenc}
\usepackage[utf8]{inputenc}
\usepackage[english]{babel}

\usepackage{latexsym}
\usepackage{graphics}
\usepackage{graphicx}
\usepackage{bm}
\usepackage{epsfig}
\usepackage{amsmath,amsfonts}
\usepackage{url}
\usepackage{epstopdf}
\usepackage{color}

\definecolor{vale}{rgb}{0,0.5, 1.}
\definecolor{david}{rgb}{1.,.5, 0.0}

\def\d{{\rm d}}

\begin{document}

\title{Cosmological constraints on the radiation released during structure formation}

\author{David Camarena \and Valerio Marra}

\institute{Departamento de Física, Universidade Federal do Esp\'{\i}rito Santo\\
Av.~F.~Ferrari, 514, 29075-910, Vit\'oria, ES, Brazil}

\date{Received: date / Revised version: date}

\abstract{
During the process of structure formation in the universe matter is converted into radiation through a variety of processes such as light from stars, infrared radiation from cosmic dust and gravitational waves from binary black holes/neutron stars and supernova explosions. The production of this astrophysical radiation background (ARB) could affect the expansion rate of the universe and the growth of perturbations. Here, we aim at understanding to which level one can constraint the ARB using future cosmological observations. We model the energy transfer from matter to radiation through an effective interaction between matter and astrophysical radiation. Using future supernova data from LSST and growth-rate data from Euclid we find that the ARB density parameter is constrained, at the 95\% confidence level, to be $\Omega_{ar_0}<0.008$. Estimates of the energy density produced by well-known astrophysical processes give roughly $\Omega_{ar_0}\sim 10^{-5}$. Therefore, we conclude that cosmological observations will only be able to constrain exotic or not-well understood sources of radiation.
}

\maketitle

\section{Introduction}
\label{sec:intro}

The process of structure formation in the universe unavoidably leads to the production of radiation.
Electromagnetic radiation is emitted during star formation, which started in small halos at redshifts of order $z\sim20$ and then peaked at $z\sim 2$~\cite{Madau:2014bja}.
Dust absorbs part of the UV radiation produced, which is then re-radiated at IR frequencies~\cite{Hauser:2001xs}.
Photons are also emitted through a variety of late-time astrophysical processes such as accreting black holes, spinning neutron stars and supernova explosions. In the latter case large amounts of relativistic neutrinos are produced~\cite{Beacom:2010kk}.

Radiation is also created in the form of gravitational waves, which are emitted both on large scales during the mergers of the supermassive black holes at the centers of galaxies and on small scales during black hole/neutron star mergers and supernova explosions.
For example, the two recent gravitational wave detections from LIGO~\cite{TheLIGOScientific:2016pea} showed that roughly 5\% of the total black hole mass has been converted into gravitational radiation.
In addition, there is also the possibility that dark matter is made of primordial black holes~\cite{Bird:2016dcv,Carr:2016drx,Sasaki:2016jop,Clesse:2016vqa} which again could radiate gravitational waves.
Finally, gravitational radiation may be produced by the backreaction of small-scale inhomogeneities on the dynamics of the background metric~\cite{Green:2010qy}.

To a first approximation one can model the radiation produced during the process of structure formation as a uniform (astrophysical) radiation background (ARB). The cosmic microwave background (CMB) -- the fossil blackbody radiation from big bang -- is not part of ARB. 
This energy transfer from matter to radiation could affect the expansion rate of the universe and the growth of perturbations. The aim of the present work is to understand to which level one can constraint the ARB using cosmological observations.
Here, we answer this question by modeling the energy transfer discussed above through an effective interaction between matter and astrophysical radiation.
We will consider future data from the Large Synoptic Survey Telescope (LSST \cite{Abell:2009aa}) and Euclid~\cite{Amendola:2016saw}; our results, therefore, should give the best possible cosmological constraints -- at least in the near future -- on the ARB.

\section{Model}
\label{sec:model}

The (twice contracted) Bianchi identity, within General Relativity, implies that the total energy-momentum tensor is conserved. Hence, a possible interaction between matter and astrophysical radiation can be modeled through an interaction current $Q^{\beta}$ which transfers energy and momentum from one source to the other with opposite direction:
\begin{equation} \label{inter}
\nabla_{\alpha} T_{m}^{\alpha \beta}= Q^{\beta}
\qquad
\nabla_{\alpha} T_{ar}^{\alpha \beta}=- Q^{\beta} \,.
\end{equation}
We will consider a phenomenological description for the effective interaction between matter and radiation.
In particular, we will consider the following simple interaction:%
\footnote{This kind of interaction (and its many variations) has been studied in order to model a possible interaction between dark matter and dark energy~(see e.g.~\cite{amendola2010dark} and references therein).}
\begin{align}
Q^{\beta} &= \Gamma  \, T_{m} \, u^{\beta}_{m} \,, \label{Q}
\end{align}
where $T_{m}=-\rho_{m}$ is the trace of the matter energy-momentum tensor so that the temporal component is  $Q^{0}  = \Gamma  \, \rho_{m}$.
For the interaction rate we will consider $\Gamma =  \alpha  \, H$, that is, we use the Hubble function $H=\dot a/a$ in order to parametrize the time dependence of the interaction.%
\footnote{The interaction here considered is formally similar to the one taking place at the end of inflation in an out-of-equilibrium decay (see~\cite{kolb1994early}, equations (5.62) and (5.67)).}
Since the energy goes from $\rho_m$ to $\rho_{ar}$ we set $\alpha >0$.
Moreover, as the production of astrophysical radiation is a recent phenomenon we will demand that $\alpha=0$ for $z \ge \bar z$ where $\bar z \sim 5\text{--}10$. In other words, the interaction is switched off at early times.
An advantage of the interaction above is that it is possible to obtain analytical solutions.

\subsection{Background}

The dynamical equations for the background are (a dot denotes a derivative with respect to cosmic time $t$):
\begin{align}
H^2 &=\frac{8\pi G}{3} \left(\rho_m + \rho_{\gamma} + \rho_{ar} \right) +\frac{\Lambda}{3} \,, \label{eq:Hz0} \\
\dot{\rho}_m  &+3H\rho_m = - \alpha  \, H \, \rho_m  \,,\label{eq:rhom0} \\ 
\dot{\rho}_{ar} &+4H\rho_{ar}= \alpha  \, H \, \rho_m \,,\label{eq:rhoar0} \\
\dot{\rho}_{\gamma} &+4H\rho_{\gamma}=0 \,,\label{eq:rhorg0}
\end{align}
where equations (\ref{eq:rhom0}-\ref{eq:rhorg0}) are the conservations equations for matter, astrophysical radiation and CMB photons, respectively, $\Lambda$ is the cosmological constant and we have assumed spatial flatness, in agreement with recent cosmological observations (see~\cite{Ade:2015xua} and references therein).

It is clear that this model is but a rough approximation to the actual process of production of radiation.
An important approximation is the use of an overall coupling constant to describe different processes which may or may not involve both dark matter and baryons.
Another approximation comes from the fact that the time dependence of the interaction rate $\Gamma$ is parametrized using the Hubble rate -- i.e.~according to a cosmological time scale -- while astrophysical processes could evolve on a shorter time scales. For example, the interaction rate could be modeled as being proportional to the matter density contrast, $\Gamma = \alpha \, \delta_m \, H_0$, as proposed in~\cite{Marra:2015iwa}.
However, the aim of this work is to understand if future observations can constrain the effective coupling $\alpha$ or, equivalently, the energy density of ARB. More precisely, we would like to know if cosmological observations can constrain astrophysical radiation to the level predicted by astrophysical models of star formation, IR background and gravitational wave production. For such a goal the simple model above should be adequate.

As we will confirm \emph{a posteriori}, for the forecasted observations we consider it is $\rho_{ar}\gg \rho_{\gamma}$. Consequently, we will neglect the well understood CMB photons from the remaining of this analysis.
For the same reason we neglect the contribution from a possibly massless neutrino.

The equations (\ref{eq:Hz0}-\ref{eq:rhoar0}) can be solved analytically:
\begin{align}
 \rho_m &= \rho_{m_0} (1+z)^{3+\alpha} \,,\label{eq:rhom} \\ 
 \rho_{ar} &= \rho_{ar_0} (1+z)^4 +\frac{\alpha}{\alpha -1} \rho_{m_0} \left[(1+z)^4 - (1+z)^{3+\alpha} \right] ,   \label{eq:rhoar} \\
& E^2(z) = \left[\Omega_{ar_0} + \frac{\alpha}{\alpha - 1} \Omega_{m_0}\right] \left( 1+z\right)^4  \nonumber\\
 & \phantom{vitoria}+\frac{1}{1-\alpha} \Omega_{m_0} (1+z)^{3+\alpha} +\Omega_{\Lambda_0} \,,\label{eq:Ez}
\end{align} 
where $\rho_{m_0}$ and $\rho_{ar_0}$ are the matter and astrophysical radiation energy densities today, respectively, 
$E(z)=H(z)/H_0$, and $\Omega_{\Lambda_0}=1-\Omega_{m_0}-\Omega_{ar_0}$.
The corresponding density parameters are:
\begin{align}
& \Omega_m(z)= \frac{ \Omega_{m_0}(1+z)^{(3+\alpha)}}{E^2(z)} \,,\label{eq:omegam} \\
& \Omega_{ar}(z) = \frac{\Omega_{ar_0}(1+z)^4 +\frac{\alpha}{\alpha -1} \Omega_{m_0}\left[(1+z)^4 -(1+z)^{3+\alpha}\right]}{E^2(z)} ,\label{eq:omegaar}
\end{align}
so that one has $ \Omega_m + \Omega_{ar} +\Omega_{\Lambda} =1$.

As discussed earlier, the energy exchange from $\rho_m$ to $\rho_{ar}$ is a recent phenomena which we model as starting at a redshift $\bar z \sim 5\text{--}10$.
Consequently, for $z \ge \bar z$ it is $\alpha=\rho_{ar}=0$.
Using this initial condition and equations \eqref{eq:rhoar} 
one then finds the present-day astrophysical density parameter as a function of initial redshift and coupling:
\begin{align}
& \Omega_{ar_0} = \left(
\frac{\alpha}{\alpha -1}\right)\left[(1+\bar z)^{\alpha -1}-1\right] \Omega_{m_0} . \label{eq:inicon}
\end{align}
As expected, $\Omega_{ar_0}$ is proportional to both the coupling parameter and the matter density.
One expects a small $\alpha$ and so $\Omega_{ar_0} \ll \Omega_{m_0}$.
For illustration purposes, the left panel of Figure~\ref{case1} shows the evolution of the background energy densities for the case $\alpha = 0.2$.

\subsection{Perturbations}

\begin{figure}
\begin{center}
\includegraphics[width= .45 \textwidth]{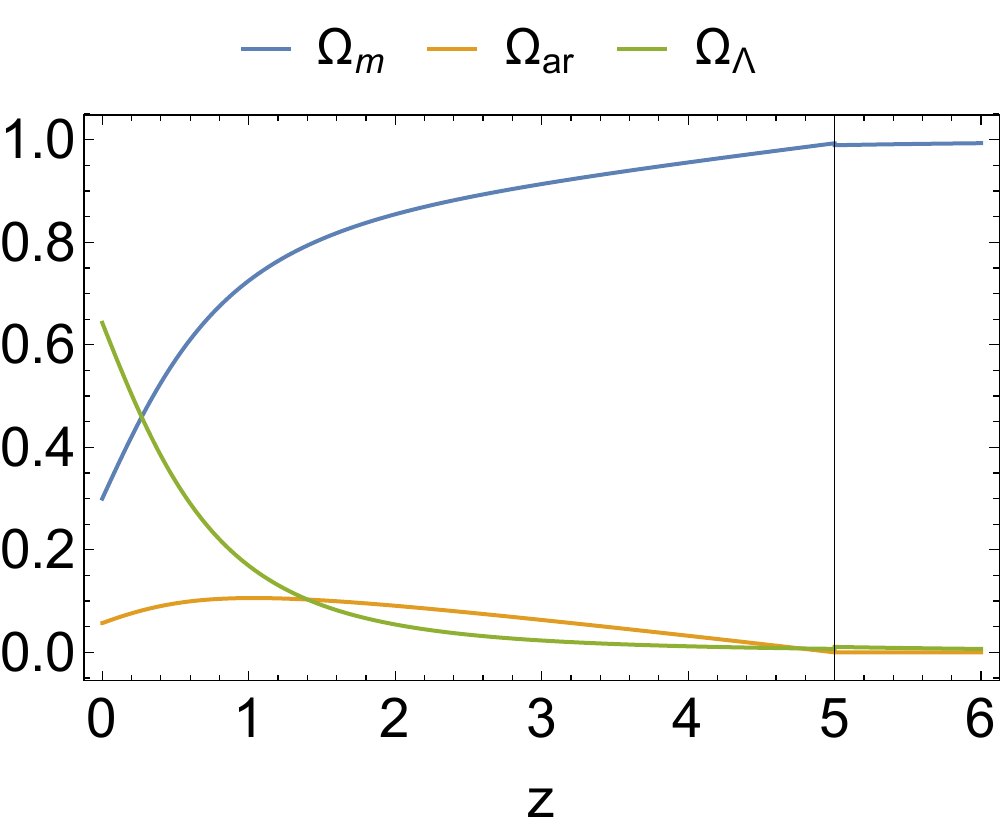}\\
\vspace{.5cm}
\includegraphics[width= .45 \textwidth]{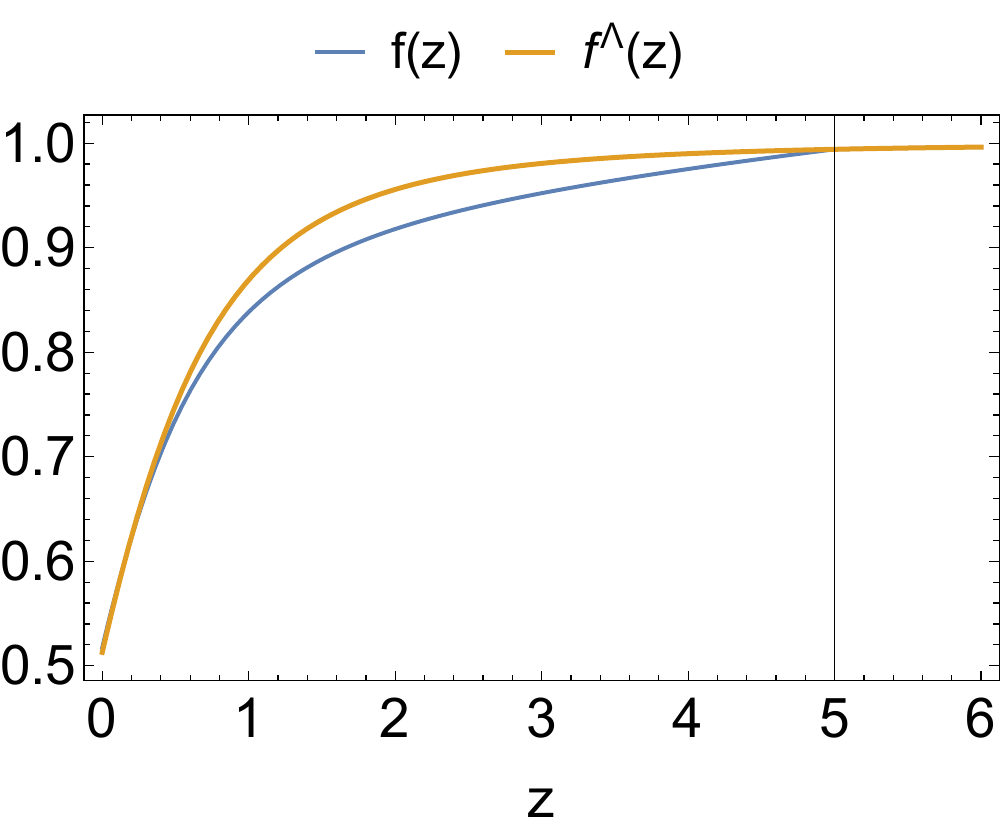}
\caption{
Top: evolution of the background energy densities for $\alpha =0.2$. At $\bar z$ the interaction is turned off and $\Omega_{ar} = 0$.
Bottom: growth rate for the model of this paper with $\alpha = 0.2$ and for the $\Lambda$CDM with the same present-day matter density.
In both plots the interaction induces larger changes for redshifts between 0 and $\bar z=5$ (vertical line). Indeed for $z\ge \bar z$ the interaction is switched off and for redshifts close to zero the cosmological constant dominates.
See Section \ref{sec:model} for more details.
}
\label{case1}
\end{center}
\end{figure}

In equation \eqref{Q} $u^{\beta}_{m}$ is the four-velocity of the matter component. As discussed in~\cite{Honorez:2010rr} this kind of interaction does not alter the Euler equation as there is no momentum transfer in the matter rest frame.
This choice should be reasonable as one does not expect a fifth force for the phenomenology discussed in this paper.
The perturbation equation for sub-horizon scales is then:
\begin{equation} \label{perteq}
\ddot \delta_m +2 H \dot \delta_m + H^2 \left (\frac{\alpha -3}{2} \right) \Omega_m \delta_m \approx 0  \,,
\end{equation}
where we made the approximation 
$\Omega_{ar} \theta_{ar}\ll \Omega_{m} \theta_m$
and $\Omega_{ar} \delta_{ar}\ll \Omega_{m} \delta_m$, and we neglected terms quadratic (or higher) in combinations of $\alpha$ and $\Omega_{ar}/\Omega_{m}$.
This means that $\theta_{\rm tot} \approx \theta_m$.
In the previous equations $\delta$ is the density contrast, $\theta= \nabla_i v^i $ is the divergence of the velocity field, and the total divergence is given by:
\begin{equation}
\theta_{\rm tot}=\frac{3\Omega_m \theta_m + 3\Omega_{ar} \theta_{ar}}{3\Omega_m +4\Omega_{ar}} \ . \label{thet}
\end{equation}
We will denote with $G(t)=\delta_m(t)/\delta_m(t_0)$ the growth function normalized to unity at the present time.
In obtaining \eqref{perteq} we have perturbed the expansion rate in $\Gamma =  \alpha  \, H$ as done in \cite{Gavela:2010tm,Li:2013bya} in order to preserve gauge invariance.
The right panel of Figure~\ref{case1} shows the evolution of the growth rate $f=\frac{\d \ln \delta_m}{\d \ln a}$ for the case $\alpha = 0.2$. For the sake of comparison, the growth rate of the $\Lambda$CDM with the same present-day matter density is also shown. 

\section{Comparison to observations}
\label{results}

\subsection{SN data}
\label{supernovae}

At the background level, we will use the forecasted supernova Ia sample relative to ten years of observations by the Large Synoptic Survey Telescope. This dataset features a total of $10^{6}$
supernovae with intrinsic dispersion of 0.12 mag in the redshift range $z=0.1-1.0$ with the redshift distribution
as given in \cite{Abell:2009aa}.

The predicted theoretical magnitudes are related to the luminosity
distance $d_{L}$ by:
\begin{equation}
m(z)=5\log_{10}\frac{d_{L}(z)}{10\,\textrm{pc}}\,,
\end{equation}
which is computed under the assumption of spatial flatness:
\begin{equation}
d_{L}(z)=(1+z)\int_{0}^{z}\frac{\d\tilde{z}}{H(\tilde{z})}\,.
\end{equation}
The $\chi^{2}$ function is then:
\begin{equation}
\chi_{\rm SNe}^{'2}=\sum_{i}\frac{[m_{i}-m(z_{i})+\xi]^{2}}{\sigma^{2}}\,,
\end{equation}
where the index $i$ labels the forecasted supernovae and $\sigma=0.12$ mag.
The parameter $\xi$ is an unknown offset sum of the supernova absolute
magnitude and other possible systematics. As
usual, we marginalize the likelihood $L'_{\rm SNe}=\exp(-\chi_{\rm SNe}^{'2}/2)$
over $\xi$, such that $L_{\rm SNe}=\int\d\xi\, L'_{\rm SNe}$, leading to a new
marginalized $\chi^{2}$ function:
\begin{equation} \label{sne}
\chi_{\rm SNe}^{2}=S_{2}-\frac{S_{1}^{2}}{S_{0}}\,,
\end{equation}
where we neglected a cosmology-independent normalizing constant, and
the auxiliary quantities $S_{n}$ are defined as:
\begin{equation}
S_{n}\equiv\sum_{i}\frac{\left[m_{i}-m(z_{i})\right]^{n}}{\sigma_{i}^{2}}\,.
\end{equation}
Note that, as $\xi$ is degenerate with $\log_{10}H_{0}$, we are effectively
marginalizing also over the Hubble constant.

\subsection{Growth rate data}
\label{fs8}

At the linear perturbation level, we will build the growth-rate likelihood using the forecasted accuracy of a future Euclid-like mission as obtained in~\cite{Amendola:2013qna}.
Growth-rate data are  given
as a set of values $d_{i}$ where
\begin{equation} \label{obsf}
d=f\sigma_{8}(z)=f(z) G(z) \sigma_{8} \,.
\end{equation}
The $\chi^{2}$ function is then:
\begin{equation} \label{rsd}
\chi_{f\sigma_{8}}^{2}=\sum_{i}\frac{[d_{i}-d(z_{i})]^{2}}{\sigma_{i}^{2}}\,,
\end{equation}
where the index $i$ labels the redshift bins which span the redshift range $0.5<z<2.1$. The uncertainties $\sigma_i$ are as given in \cite{Amendola:2013qna} (Table II).

\subsection{Planck prior}

The $\chi^{2}$ functions above depend on three parameters: $\alpha$, $\Omega_{m_0}$ and $\sigma_8$. It is useful to consider a prior on the latter two parameters in order to reduce possible degeneracies.
As the interaction we are considering is absent at earlier times -- and so the cosmology is unchanged for $z\ge \bar z$ -- we can use a prior on $\Omega_{m_0}$ and $\sigma_8$ from the CMB.
However, as the evolution for $z< \bar z$ is different, we have to adopt effective present-day parameters in building the prior. Specifically, by demanding $\Omega_m^{\Lambda}(\bar z) = \Omega_m(\bar z)$ and $\sigma^{\Lambda}_8(\bar z) =\sigma_8(\bar z)$ we find that the effective parameters that we have to use are:
\begin{align}
\Omega_{m_0}^{\Lambda} &=\Omega_{m_0}(1+\bar z)^{\alpha} \left[\frac{E^{\Lambda}(\bar z)}{E(\bar z)}\right]^2 \,,\label{omegaprior} \\
\sigma_{8}^{\Lambda} &= \sigma_{8} \frac{G^{\Lambda}(\bar z)}{G(\bar z)} \,,\label{sigmaprior}
\end{align}
where $E^{\Lambda}(z)$ and $G^{\Lambda}(z)$ are the corresponding functions in the $\Lambda$CDM case (i.e.~with $\alpha=0$).

From Figure~19 (TT, TE, EE+lowP) of Planck 2015 XIII~\cite{Ade:2015xua} one can deduce the covariance matrix between $\Omega_{m_0}$ and $\sigma_8$ (we approximate the posterior as Gaussian):  $\sigma_{\Omega_m}=0.009$, $\sigma_{\sigma_8}=0.014$ and $\rho \simeq 0$.
Consequently, the $\chi^{2}$ function of the CMB prior is:
\begin{equation}
\chi_{cmb}^{2}=  \label{cmbprior}
\left[\frac{\Omega_{m_0}^{\rm fid} -\Omega_{m_0}^{\Lambda}}{\sigma_{\Omega_m}}\right]^2 + 
\left[ \frac{\sigma_{8}^{\rm fid} -\sigma_{8}^{\Lambda}} {\sigma_{\sigma_8}}\right]^2  \,.
\end{equation}
%

\subsection{Full likelihood}

The full likelihood is based on the total $\chi^{2}$ which is:
\begin{equation}
\chi_{{\rm tot}}^{2}=\chi_{SNe}^{2}+\chi_{f\sigma_{8}}^{2} + \chi_{cmb}^{2}\,. \label{chi2tot}
\end{equation}
Furthermore, since energy is transferred from $\rho_m$ to $\rho_{ar}$, we adopt the following flat prior on the coupling constant: $\alpha \ge 0$.
Our fiducial model is specified by the following values of the parameters: $\alpha=0$, $\Omega_{m0}=0.3$ and $\sigma_8=0.8$.

The datasets we consider should give the tightest constraints -- at least for the near future -- as far as background and perturbation observables are concerned. One way to improve the results of the next section could be to place a low-redshift prior on $\sigma_8$, for which one has to extend the theory of nonlinear structure formation (mass function, bias, etc) to the case of the interaction here considered.

\section{Results}
\label{sec:results}

\begin{figure*}
\begin{center}
\includegraphics[width= .8 \textwidth]{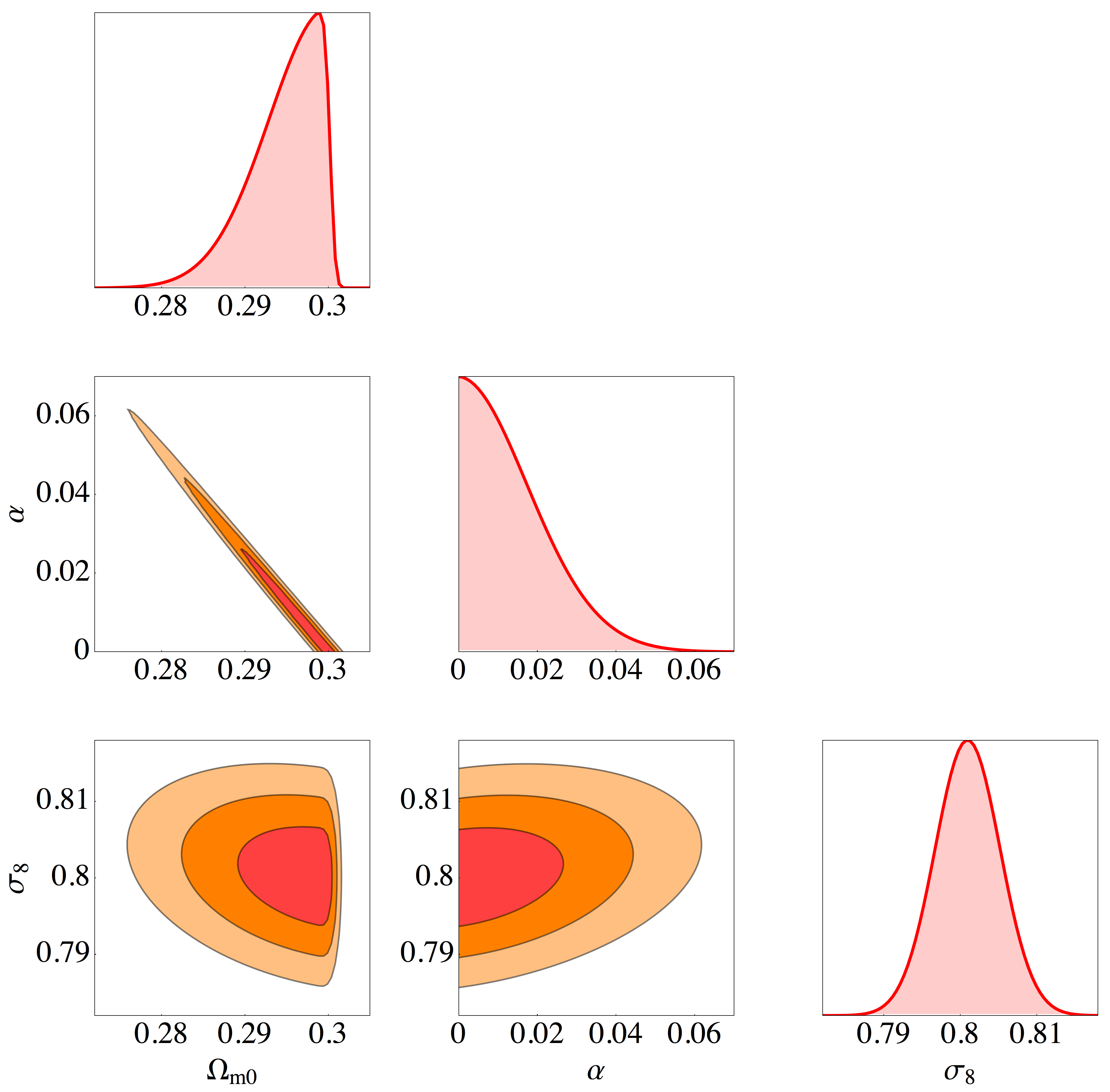}
\caption{
Marginalized 1-, 2- and 3$\sigma$ constraints and correlations on the parameters $\alpha$, $\Omega_{m0}$ and $\sigma_8$ using the total likelihood built from the $\chi^2$ function of equation~\eqref{chi2tot}.
The starting redshift of $\bar z =5$ has been adopted. 
See Section \ref{sec:results} for more details.
}
\label{constraints}
\end{center}
\end{figure*}

\begin{figure}
\begin{center}
\includegraphics[width= .45 \textwidth]{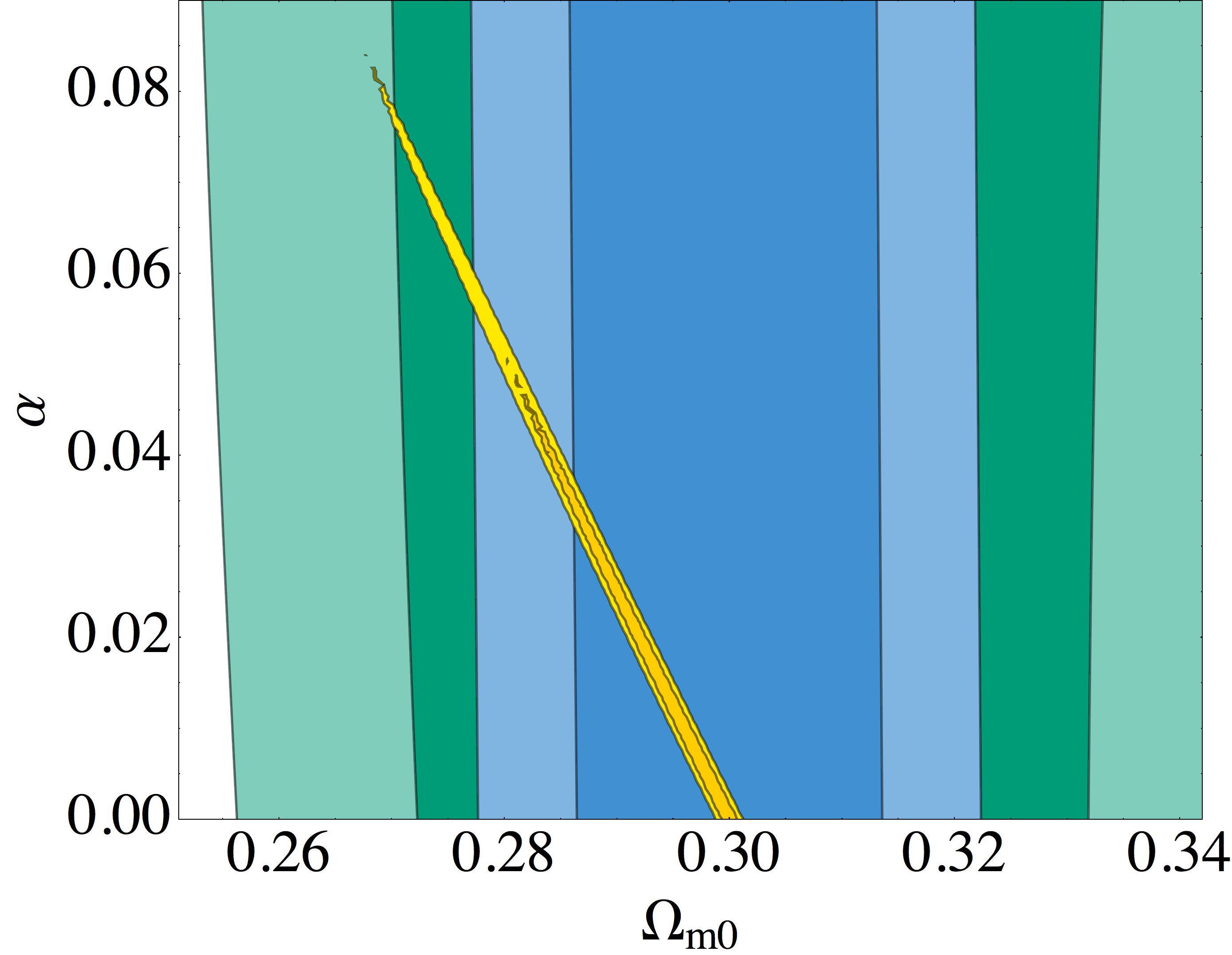}\\
\vspace{.5cm}
\includegraphics[width= .45 \textwidth]{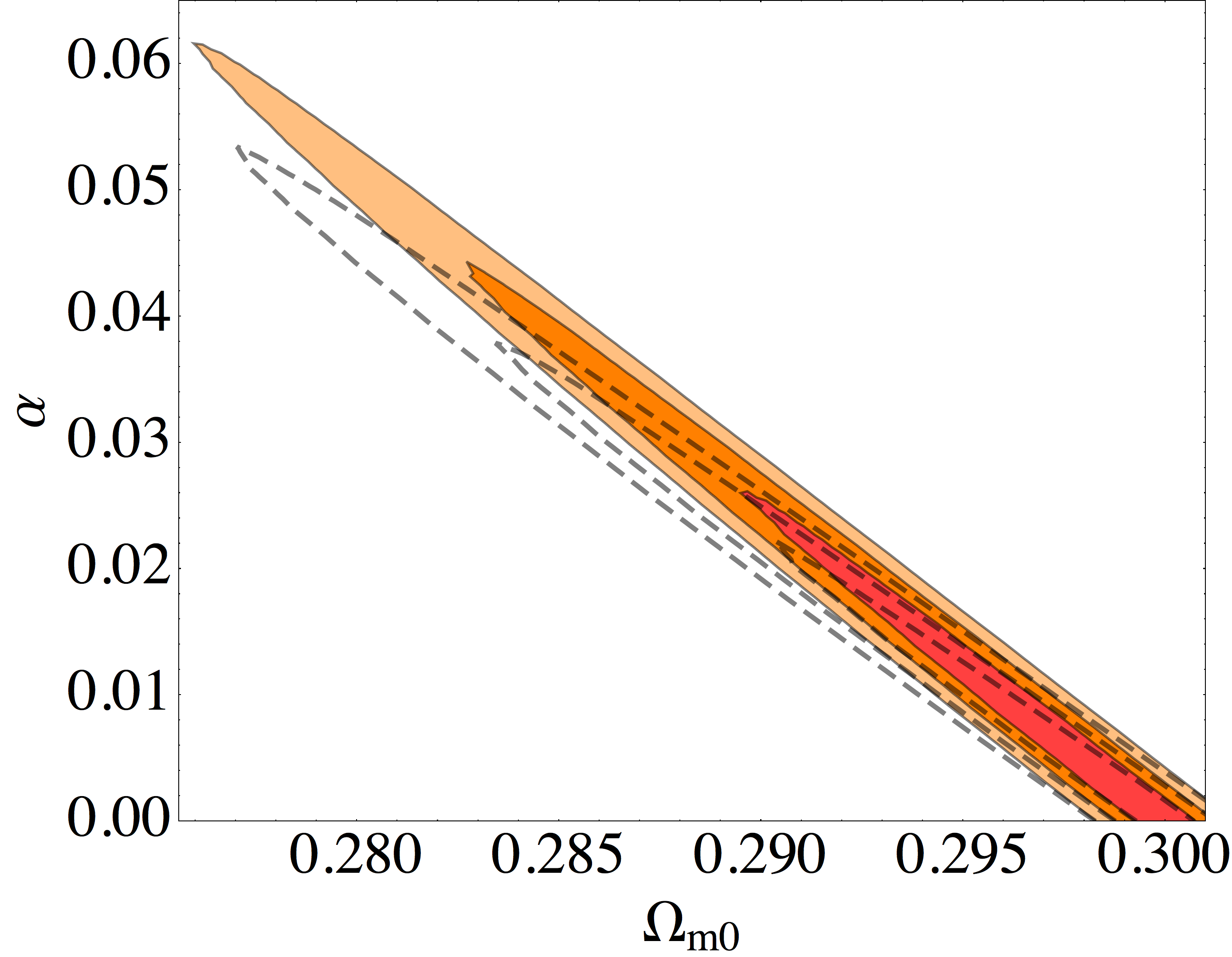}
\caption{
Top: 
marginalized 1- and 2$\sigma$ constraints on $\alpha$ and $\Omega_{m0}$ for the SN likelihood of \eqref{sne} (yellow contours), the growth rate likelihood of \eqref{rsd} (green contours) and the CMB prior of \eqref{cmbprior}  (blue contours).
The combination of these constraints give the corresponding panel of Figure~\ref{constraints}.
Bottom: 
the dashed empty contours show the constraints on $\alpha$ and $\Omega_{m0}$ for the case $\bar z = 10$. For comparison-sake, the red-to-orange contours from Figure~\ref{constraints} corresponding to the case $\bar z =5$ are also shown.
See Section \ref{sec:results} for more details.
}
\label{extra}
\end{center}
\end{figure}

Figure~\ref{constraints} shows marginalized 1-, 2- and 3$\sigma$ constraints and correlations on the parameters $\alpha$, $\Omega_{m_0}$ and $\sigma_8$ using the total likelihood built from the $\chi^2$ function of equation~\eqref{chi2tot}.
The left panel of Figure~\ref{extra} shows the marginalized 1- and 2$\sigma$ constraints on $\alpha$ and $\Omega_{m_0}$ for each of the three individual likelihoods (LSST supernovae, Euclid growth-rate data and Planck prior). It is clear that the strong degeneracy between $\alpha$ and $\Omega_{m_0}$  comes from the supernova likelihood, and that growth data and CMB prior marginally help at constraining $\Omega_{m_0}$.
At 95\% confidence level we find that $\alpha < 0.035$.

In the analysis of Figures~\ref{constraints} and~\ref{extra} (left panel) the starting redshift of $\bar z =5$ has been adopted. In the right panel of Figure~\ref{extra} we present constraints on $\alpha$ and $\Omega_{m_0}$ for the case $\bar z =10$.
The 95\% confidence level constraint on the coupling is now slightly tighter: $\alpha < 0.03$.
As one can see, the results do not depend strongly on $\bar z$.
As the processes contributing to the ARB start and take place at different redshifts, the fact that our results do not depend strongly on $\bar z$ means that our modeling and approximations are consistent.

\section{Discussion}
\label{sec:discussion}

In order to make contact with astrophysical bounds on the ARB it is useful to express our results not with respect to the coupling constant $\alpha$ but with respect to the present-day density of astrophysical radiation $\Omega_{ar_0}$. Using equation \eqref{eq:inicon} it is easy to make this change of variable and obtain from the results of the previous Section that, at the 95\% confidence level, it is $\Omega_{ar_0}<0.008$. The latter constraint depends weakly on $\bar z$.

The total extra galactic background light (EBL) has a density parameter of the order of $\Omega_{ebl} \sim 5 \cdot 10^{-6}$, roughly a factor 10 times smaller than the density parameter relative to the CMB photons~\cite{Hauser:2001xs}.

Regarding the stochastic gravitational-wave background, one usually defines the density parameter $\Omega_{gw}(f)$ within the logarithmic frequency interval between $f$ and $f+\d f$.
For $f<100$Hz the spectrum is well approximated by a power law, $\Omega_{gw}(f)\propto f^{2/3}$, while for $f>100$Hz it quickly drops. Using the normalization $\Omega_{gw} (f=25 \text{Hz}) \sim 10^{-9}$ \cite{TheLIGOScientific:2016wyq} one can then integrate $\Omega_{gw}(f)$ and obtain $\Omega_{gw}\sim 5 \cdot 10^{-9}$.
The latter estimate only considers gravitational waves from binary black holes. Therefore, the total energy density in the stochastic gravitational-wave background is somewhat larger, see~\cite{Rosado:2011kv,2016arXiv160806889R} for a comprehensive reviews.
If a fraction of dark matter is made of primordial black holes, one would expect an additional gravitational wave background coming from their stochastic mergers~\cite{Bird:2016dcv,Carr:2016drx,Sasaki:2016jop}. However, this background is supposed to be subdominant as compared to the one produced by black holes which were the result of star formation and evolution~\cite{Mandic:2016lcn}. Moreover, the merger rate of primordial black holes is not negligible in the past, contrary to our assumption that $\alpha=0$ for $z \ge \bar z$.

A sizable contribution to the ARB comes from the diffuse supernova neutrino background (DSNB). The cosmic energy density in neutrinos from core-collapse supernovae is expected to be comparable to that in photons from stars~\cite{Beacom:2010kk}.
Indeed, one single core-collapse supernova produces $\sim 3 \cdot 10^{53}$ erg in MeV neutrinos and such supernovae occur approximately every 100 years in the Milky Way.
One can then conclude that the average neutrino power of our Galaxy is $\sim 10^{44}$~erg/s, similar to its IR-optical luminosity.%
\footnote{Note that the total energy density of the high-energy (>100 TeV) neutrino background recently measured by IceCube is negligible as compared to the DSNB~\cite{Aartsen:2013jdh}.}
One can then estimate that $\Omega_{dsnb} \sim 5 \cdot 10^{-6}$.

Finally, as discussed in the Introduction, backreaction could contribute to the ARB at not well understood rates~\cite{Green:2010qy}.
Summing up, one expects the ARB to have a density parameter of the order of $\Omega_{ar_0}\sim 10^{-5}$.

\section{Conclusions}
\label{sec:conclusions}

We have computed how well future cosmological observations can constrain the energy density of the astrophysical radiation background (ARB). ARB is the (to a first approximation) uniform radiation density produced during the process of structure formation in the recent universe.
We modeled the energy transfer from matter to radiation through an effective interaction between matter and astrophysical radiation, which is set to be zero at a redshift of about 5--10.
Using forecasted supernovas from the LSST and growth rate data from Euclid we found that the coupling constant $\alpha$ is constrained to be $\alpha < 0.035$ and the present-day density of astrophysical radiation to be $\Omega_{ar_0}<0.008$ (both at the 95\% confidence level).

Estimates of the energy density produced by well-known astrophysical processes give roughly $\Omega_{ar_0}\sim 10^{-5}$,
almost three orders of magnitude smaller than the upper limit that can be obtained with LSST and Euclid.
Therefore, we conclude that cosmological observations will be able to constrain only exotic not-well understood sources of radiation such as the backreaction of small-scale inhomogeneities on the dynamics of the universe.

\section{Acknowledgments}

It is a pleasure to thank Júlio Fabris, Oliver Piattella, Davi Rodrigues and Winfried Zimdahl for useful discussions.
DCT is supported by the Brazilian research agencies CAPES.
VM is supported by the Brazilian research agency CNPq.

\bibliographystyle{utcaps}
\bibliography{refs}

\end{document}